\documentclass[aps,prd,twocolumn,numerical,superscriptaddress,floatfix,nofootinbib]{revtex4}

\usepackage{float,ulem}
\usepackage{graphicx,epsfig,epstopdf,amsmath,amssymb}
\usepackage{xcolor}
\usepackage[colorlinks=true,linkcolor=blue,citecolor=blue,urlcolor=
blue]{hyperref}

\newcommand{\be}{\begin{equation}}
\newcommand{\ee}{\end{equation}}
\newcommand{\ba}{\begin{eqnarray}}
\newcommand{\ea}{\end{eqnarray}}
\newcommand{\nn}{\nonumber\\}

\usepackage{caption}
\usepackage{subcaption}

\begin{document}
\title{Is first-order relativistic hydrodynamics in general frame stable and causal for arbitrary interaction?}

\author{Rajesh Biswas}
\email{rajeshbiswas@niser.ac.in}
\affiliation{School of Physical Sciences, National Institute of Science Education and Research, HBNI, 752050, Jatni, India.}

\author{Sukanya Mitra}
\email{sukanya.mitra10@gmail.com}
\affiliation{Department of Nuclear and Atomic Physics, Tata Institute of Fundamental Research, Homi Bhabha Road, Mumbai 400005, India.}

\author{Victor Roy}
\email{victor@niser.ac.in}
\affiliation{School of Physical Sciences, National Institute of Science Education and Research, HBNI, 752050, Jatni, India.}

\begin{abstract}
We derive a first-order, stable and causal, relativistic hydrodynamic theory from the microscopic kinetic equation using the gradient expansion technique in a general frame. The general frame is introduced from 
the arbitrary matching conditions for hydrodynamic fields. The interaction is introduced in the relativistic Boltzmann equation through the momentum-dependent relaxation time approximation (MDRTA) 
with the proposed collision operator that preserves the conservation laws. We demonstrate here for the first time that not only the general frame choice, but also the momentum dependence of microscopic 
interaction rate, captured through MDRTA, is imperative for producing the essential field corrections that give rise to a causal and stable first-order relativistic theory.
\end{abstract}
\maketitle

\textit{Introduction.}$\textendash$
The hydrodynamic theory is an effective coarse-grained formulation of the underlying microscopic dynamics at the long-wavelength limit, that has served for decades as an efficient and accessible tool for a vast 
range of problems in theoretical physics. However convenient, the relativistic extension of the first-order dissipative Navier-Stokes (NS) formalism introduced by Landau-Lifshitz (LL)~\cite{LL} and Eckart
~\cite{Eckart}, encounters severe issues with instability~\cite{Hiscock:1983zz,Hiscock:1985zz,Hiscock:1987zz} and superluminal signal propagation, which pose serious limitation to the practical application of the 
theory. Later on, second-order Muller-Israel-Stewart (MIS) theory~\cite{Muller:1967zza,Israel:1976tn,Israel:1979wp}
and some of its extended versions ~\cite{Muronga:2001zk,Denicol:2012cn,Baier:2007ix,Jaiswal:2013vta,Grozdanov:2015kqa} are introduced to 
remedy these problems. Recently, a new study has been proposed by Bemfica, Disconzi, Noronha, and Kovtun (BDNK)~\cite{Bemfica:2017wps,Bemfica:2019knx,Bemfica:2020zjp,Kovtun:2019hdm,Hoult:2020eho,Hoult:2021gnb,
Rocha:2022ind} for a first-order stable and causal theory by defining the out of equilibrium hydrodynamic variables in a general frame other than LL or Eckart through their postulated constitutive relations that 
include both time and space gradients.

In this work, we have derived a first-order theory using gradient expansion technique in an arbitrary frame where the explicit expressions of the field redefinition coefficients have been estimated from the
underlying microscopic dynamics. The homogeneous part of the out of equilibrium momentum distribution has been extracted from the hydrodynamic matching conditions. The inhomogeneous part obtained from the 
Boltzmann equation becomes sensitive to the system interactions through its collision term. The relaxation time approximation (RTA)~\cite{Relax} is proven to be a convenient form for linearization of the 
collision kernel with a wide range of applications (see Ref.~\cite{Florkowski:2017olj} and references therein) and its momentum dependence can be related to the microscopic interaction relevant 
for the medium under consideration \cite{Dusling:2009df}. These two facts provide a strong motivation to use momentum dependent relaxation time approximation (MDRTA) in the relativistic transport equation to 
obtain the inhomogeneous part of the solution ~\cite{Teaney:2013gca,Kurkela:2017xis,Rocha:2021lze,Rocha:2021zcw,Mitra:2020gdk,Mitra:2021owk,Mitra:2021ubx,Dash:2021ibx}. Here we 
propose a new collision operator under MDRTA which obeys the fundamental microscopic and macroscopic conservation laws irrespective of particular momentum dependence of RTA or the matching indices. With this formalism, here we have analytically calculated  the values of the coefficients in the constitutive relations of hydrodynamic field redefinition from the kinetic theory in a general frame, 
i.e., for arbitrary matching conditions.

We further analyse the dispersion relation resulting from small perturbations around the hydrostatic equilibrium for this first-order theory to investigate the stability and causality of the system. It is 
observed that the first-order field correction coefficients responsible for generating causal and stable modes are directly related to the microscopic dynamics of the system. Even in a general frame where the 
first-order theory is expected to be causal and stable, we find that only non-zero momentum dependence of relaxation time gives rise to the causal and stable modes. The stability and causality conditions critically 
depend upon the particular momentum dependence of MDRTA. These are the key findings of the current work. To the best of our knowledge, for the first time, a correlation between the interaction dynamics and the 
causality and stability of a relativistic fluid is being reported.
 
Throughout the manuscript, we have used natural unit~($\hbar = c = k_{B} = 1 $) and flat space-time with mostly negative metric $g^{\mu\nu} = \text{diag}\left(1,-1,-1,-1\right)$. 

\textit{Hydrodynamic field redefinition.}$\textendash$
The basic idea is to employ the relativistic Boltzmann transport equation to estimate the out-of-equilibrium one-particle distribution function $f(x,p)$ for general hydrodynamic frame (defined later),
\be
p^{\mu}\partial_{\mu}f(x,p)=C[f]=-\cal{L}[\phi]~.
\label{RTE}
\ee
Here $p$ is the particle four-momenta and  $x$ denotes the space-time variable, $f=f^{(0)}+f^{(0)}(1\pm f^{(0)})\phi$ with $f^{(0)} (=[exp(\frac{p\cdot u}{T}-\frac{\mu}{T})\mp 1]^{-1}$ for Bosons and Fermions 
respectively) as the equilibrium distribution and $\phi$ is the out of equilibrium deviation; $C[f]$ is the collision integral corresponds to the two-to-two elastic collisions. It is linearized as 
${\cal{L}}[\phi]=\int d\Gamma_{p_1}d\Gamma_{p'}d\Gamma_{p'_1}f^{(0)}f_1^{(0)}(1\pm f'^{(0)})(1\pm f_{1}'^{(0)})\{\phi+\phi_1-\phi'-\phi'_1\}W(p'p'_1|pp_1)$, with $d\Gamma_p=\frac{d^3p}{(2\pi)^3p^0}$ and $W$ is the 
transition rate that depends on the cross-section of the interactions. In gradient expansion technique $\phi$ is expressed as $\phi=\sum_{r}\phi^{(r)}$, with $\phi^{(r)}$ as the  $r^{th}$ order  out-of-equilibrium 
deviation of the distribution function.

In general, $\phi^{(r)}$ can be expressed as a linear combination of $r^{th}$ order field gradients with appropriate tensor coefficients~\cite{Degroot}:
\be
\phi^{(r)}=\sum_{l}A^{(r)}_{l} X^{(r)l}+\sum_{m} B_{m}^{(r)\mu}Y^{(r)m}_{\mu}+\sum_{n} C_{n}^{(r)\mu\nu}Z^{(r)n}_{\mu\nu}~,
\label{phi1}
\ee
with $X^{(r)l}, Y^{(r)m}_{\mu}$ and $Z^{(r)n}_{\mu\nu}$ are the $r^{th}$ order scalar, vector and rank-2 tensor gradient corrections of $l, m$, and $n^{th}$ kind respectively. $A_{l}^{(r)}, B_{m}^{(r)\mu}$ and 
$C_{n}^{(r)\mu\nu}$ are the unknown coefficients functions of space-time, particle momentum and the ratio of it's rest mass to the temperature $z=m/T$. We expand the coefficients in a polynomial basis to extract 
their values as, $A_{l}^{(r)}=\sum_{s=0}^{\infty}A_{l}^{r,s}(z, x) P_s^{(0)}, B_{m}^{(r) \mu}=\sum_{s=0}^{\infty} B_{m}^{r,s}(z, x) P_s^{(1)} \tilde{p}^{\langle \mu \rangle}$, 
$C_{n}^{(r) \mu \nu}=\sum_{s=0}^{\infty} C_{n}^{r,s}(z, x)P_s^{(2)} \tilde{p}^{\langle \mu} \tilde{p}^{\nu \rangle}$. Inspired from ~\cite{Dobado:2011qu} and being convenient for the current analysis, we employ 
an orthogonal polynomial basis which are partially orthogonal in the scalar sector. For our case the first two polynomials, $P^{(0)}_0=1,~P^{(0)}_1=\tilde{E}_p$ are not orthogonal but all other higher polynomials 
are chosen to be orthogonal to these two as well as among themselves and monic (in $P^{(n)}_s$ the coefficient of maximum power of $\tilde{E}_p$, i.e, ${\tilde{E}_p}^s$ is 1). Concisely they are given by,
\begin{align}
&P^{(0)}_0=1~,~~P^{(0)}_1=\tilde{E}_p~,~~P^{(1)}_0=1~,~~P^{(2)}_0=1~,
\label{poly1}\\
&\int dF_p(\tilde{E}_p/\tau_R)(\Delta_{\mu\nu}p^{\mu}p^{\nu})^n P_s^{(n)}P_r^{(n)}\sim\delta_{s,r}~,
\label{poly2}
\end{align}
with $\tau_R$ as the relaxation time of single particle distribution function that will be introduced later with more details. 
The used notations are, $ dF_p=d\Gamma_pf^{(0)}(1\pm f^{(0)}),~\tilde{p}^{\mu}=p^{\mu}/T$, $\tilde{\mu}=\mu/T$, $\tilde{E}_{p} = u_{\mu}p^{\mu}/T,~ \tilde{p}_{\langle\mu\rangle}=\Delta_{\mu\nu}\tilde{p}^{\nu}$ and 
$\tilde{p}_{\langle\mu} \tilde{p}_{\nu\rangle} =\Delta_{\mu\nu}^{\alpha\beta}\tilde{p}_{\alpha}\tilde{p}_{\beta}$  with $T,\mu$ and $u^{\mu}$ as the temperature, chemical potential and fluid four velocity of the 
system at equilibrium and $\Delta^{\mu\nu}=g^{\mu\nu}-u^{\mu}u^{\nu}$.

It is observed that, by the virtue of the collision integral properties ${\cal{L}}[1]=0$ and ${\cal{L}}[p^{\mu}]=0$, that follow from the particle number and energy-momentum conservation respectively, 
the coefficients $A_{l}^{r,0}, A_{l}^{r,1}$ and $B_{m}^{r,0}$ can not be determined from the transport equation~\eqref{RTE} and hence they are called the coefficients of the homogeneous solution. The rest 
of the coefficients $A_{l}^{r,s}, B_{m}^{r,s}$ and $C_{n}^{r,s}$ can be estimated from the transport equation and they are called inhomogeneous or interaction solutions.  We take the recourse of the matching 
conditions which are constraints that set the thermodynamic fields, such as temperature, chemical potential, etc., to their equilibrium values even in the presence of dissipation,  to extract the coefficients
of the homogeneous part of the distribution function. Each such matching conditions produce one out of an infinite number of possible ``hydrodynamic frames''~\cite{Hoult:2021gnb}. From the requirement of 
setting two scalars and one vector homogeneous coefficients from these constraints, we use the following three matching conditions,
\be
\int dF_p\tilde{E}_{p}^{i}\phi=0,~\int dF_p\tilde{E}_{p}^{j}\phi=0,~\int dF_p\tilde{E}_{p}^{k}\tilde{p}^{\langle\mu\rangle}\phi=0,
\label{match}
\ee
with $i\neq j$, $i,j,k$ are non-negative integers. We identify the set of matching indices $(1,2,1)$ and $(1,2,0)$ to represent the LL and Eckart frame respectively. Substituting Eq.~\eqref{phi1} in Eq.~\eqref{match}, 
we find the homogeneous part in terms of the interaction part 
$\phi_{\text{int}}^{(r)}=\sum_{s=2}^{\infty}P_s^{(0)}\sum_{l}A^{r,s}_{l}X^{(r)l}+\tilde{p}^{\langle\mu\rangle}\sum_{s=1}^{\infty}P^{(1)}_s\sum_{m}B^{r,s}_{m}Y^{(r)m}_{\mu}+
\tilde{p}^{\langle\mu}\tilde{p}^{\mu\rangle}\sum_{s=0}^{\infty}P^{(2)}_s \sum_{n}C^{r,s}_{n} Z^{(r)n}_{\mu\nu}$ and the matching indices.
Using this prescription, the entire out of equilibrium distribution function for any order becomes,
\begin{align}
\phi^{(r)}=\phi^{(r)}_{\text{int}}
&-\tilde{E}_{p}\left[\frac{I_{j}}{\mathcal{D}_{i,j}^{1,0}}\int dF_p\tilde{E}_{p}^{i}\phi_{\text{int}}^{(r)}+(i\leftrightarrow j)\right]\nn
&-\left[\frac{I_{j+1}}{\mathcal{D}_{i,j}^{0,1}}\int dF_p\tilde{E}_{p}^{i}\phi_{\text{int}}^{(r)}+(i\leftrightarrow j)\right]\nn
&-\frac{\tilde{p}_{\langle\nu\rangle}}{J_k}\int dF_p\tilde{E}_{p}^{k} \tilde{p}^{\langle\nu\rangle}\phi_{\text{int}}^{(r)}~.
\label{phi5}
\end{align}
Here we use the shorthand notation: $\mathcal{D}_{i,j}^{m,n}=I_{i+m}I_{j+n}-I_{i+n}I_{j+m}$ with the properties 
$\mathcal{D}_{i,j}^{m,n} = - \mathcal{D}_{i,j}^{n,m}$ and $\mathcal{D}_{i,j}^{m,n} = - \mathcal{D}_{j,i}^{m,n}$. The moment integrals are defined as 
$I_{n}=\int dF_p \tilde{E}_{p}^n,~\Delta^{\mu\nu}J_n=\int dF_p \tilde{p}^{\langle\mu\rangle}\tilde{p}^{\langle\nu\rangle}\tilde{E}_{p}^n$.
Eq.~\eqref{phi5} provides the out-of-equilibrium parts of the two most general hydrodynamic field variables, namely the particle four-flow ($N^{\mu}$) and the energy-momentum tensor ($T^{\mu\nu}$) respectively 
for the $r^{th}$ order of gradient correction as,
\be
\delta N^{(r)\mu}=\int dF_p p^{\mu}\phi^{(r)},~~\delta T^{(r)\mu\nu}=\int dF_p p^{\mu}p^{\nu}\phi^{(r)}~.
\label{field}
\ee
Utilizing Eq.~\eqref{field}, the non-equilibrium correction to the particle number density ($\delta n^{(r)}=u^{\mu}\delta N^{(r)}_{\mu}$), the energy density 
($\delta \epsilon^{(r)}=u^{\mu}u^{\nu}\delta T_{\mu\nu}^{(r)}$), pressure ($\delta P^{(r)}=-\frac{1}{3}\Delta^{\mu\nu}\delta T_{\mu\nu}^{(r)}$), energy flux or momentum density  
($W^{(r)\alpha}=\Delta^{\alpha}_{\mu}u_{\nu}\delta T^{(r)\mu\nu}$), and the particle flux ($V^{(r)\alpha}=\Delta^{\alpha}_{\mu}\delta N^{(r)\mu}$) can be estimated order-by-order as, 
\begin{align}
\delta n^{(r)}&=\int dF_p(p^{\mu}u_{\mu})\phi^{(r)}_{\text{int}}+a^{(r)} \frac{\partial n_0}{\partial\tilde{\mu}}+ u^{\mu}b^{(r)}_{\mu} T\frac{\partial n_0}{\partial T},
\label{rho2}\\
\delta\epsilon^{(r)}&=\int dF_p(p^{\mu}u_{\mu})^2\phi^{(r)}_{\text{int}}+a^{(r)} \frac{\partial\epsilon_0}{\partial\tilde{\mu}}+ u^{\mu}b^{(r)}_{\mu} T\frac{\partial\epsilon_0}{\partial T}~,
\label{en2}\\
\delta P^{(r)}&=\frac{1}{3}\int dF_p\big\{(p^{\mu}u_{\mu})^2-m^2\big\}\phi^{(r)}_{\text{int}}+a^{(r)} \frac{\partial P_0}{\partial\tilde{\mu}}
\nn
&\hspace*{01cm}+ u^{\mu}b^{(r)}_{\mu} T\frac{\partial P_0}{\partial T}~,
\label{Pr2}
\\
W^{(r)\mu}&=\int dF_p p^{\langle\mu\rangle}(p^{\mu}u_{\mu})\phi^{(r)}_{\text{int}}-(\epsilon_0+P_0)\Delta^{\mu\nu}b^{(r)}_{\nu},
\label{momflux2}\\
V^{(r)\mu}&=\int dF_p p^{\langle\mu\rangle}\phi^{(r)}_{\text{int}}-n_0\Delta^{\mu\nu}b^{(r)}_{\nu}~.
\label{pnflux2}
\end{align}
Here, $n_0, \epsilon_0$, and $P_0$ are the equilibrium values of particle number density, energy density, and pressure, respectively. $a^{(r)}$ and $b^{(r)\mu}$ are the dimensionless momentum independent quantities 
given by, $a^{(r)}=\frac{I_{i+1}}{\mathcal{D}_{i,j}^{0,1}}\int dF_p\tilde{E}_{p}^j\phi_{\text{int}}^{(r)}+(i\leftrightarrow j)~,
u^{\mu}b_{\mu}^{(r)}=\frac{I_i}{\mathcal{D}_{i,j}^{1,0}}\int dF_p\tilde{E}_{p}^j\phi_{\text{int}}^{(r)}+(i\leftrightarrow j)~,
\Delta^{\mu\nu}b_{\nu}^{(r)}=-\frac{1}{J_k}\int dF_p\tilde{E}_{p}^k \tilde{p}^{\langle\mu\rangle}\phi_{\text{int}}^{(r)}~,$ that define the homogeneous part of 
$\phi$ as, $\phi^{(r)}_{\text{h}}=a^{(r)}+b^{(r)\mu}\tilde{p}_{\mu}$. Adding up the field corrections for all orders, the most general expressions for $N^{\mu}$ and $T^{\mu\nu}$ are given by,
\begin{align}
N^{\mu}=&(n_0+\delta n)u^{\mu}+V^{\mu}~,
\label{numberflow}\\
T^{\mu\nu}=&\left(\epsilon_0+\delta\epsilon\right)u^{\mu}u^{\nu}-\left(P_0+\delta P\right)\Delta^{\mu\nu}\nn
&+\left(W^{\mu}u^{\nu}+W^{\nu}u^{\mu}\right)+\pi^{\mu\nu}~,
\label{enmomflow}
\end{align}
with $\pi^{\mu\nu}$ as the shear stress tensor.

\textit{First-order theory with MDRTA.}$\textendash$
Up to now, the discussion was completely general, and the results are applicable for any order in the gradient expansion. To provide the explicit expression for the distribution function from Eq.~\eqref{phi5}, 
one needs to estimate the interaction part of the distribution function for a specific order. For this purpose, we employ here the momentum-dependent relaxation time approximation (MDRTA) for solving 
the relativistic transport equation~\eqref{RTE} as a dynamical model study. The idea is to replace ${\cal{L}}[\phi]$ in Eq.~\eqref{RTE} with the Anderson-Witting type relaxation kernel, but now we generalise the 
relaxation time to be momentum dependent. For this purpose, we propose here a collision operator under MDRTA in Eq.\eqref{RTE} as the following, 
\begin{align}
{\cal{L}}_{\text{MDRTA}}[\phi]
&=\frac{\left(p\cdot u\right)}{\tau_R}f^{(0)}(1\pm f^{(0)})\bigg[\phi\nonumber\\
&-\frac{\langle\frac{\tilde{E}_p}{\tau_R}\tilde{E}_p^2\rangle\langle\frac{\tilde{E}_p}{\tau_R}\phi\rangle-\langle\frac{\tilde{E}_p}{\tau_R}\tilde{E}_p\rangle\langle\frac{\tilde{E}_p}{\tau_R}\phi\tilde{E}_p\rangle}
{\langle\frac{\tilde{E}_p}{\tau_R}\rangle\langle\frac{\tilde{E}_p}{\tau_R}\tilde{E}_p^2\rangle-\langle\frac{\tilde{E}_p}{\tau_R}\tilde{E}_p\rangle^2}\nonumber\\
&-\tilde{E}_p\frac{\langle\frac{\tilde{E}_p}{\tau_R}\tilde{E}_p\rangle\langle\frac{\tilde{E}_p}{\tau_R}\phi\rangle-\langle\frac{\tilde{E}_p}{\tau_R}\rangle\langle\frac{\tilde{E}_p}{\tau_R}\phi\tilde{E}_p\rangle}
{\langle\frac{\tilde{E}_p}{\tau_R}\tilde{E}_p\rangle^2-\langle\frac{\tilde{E}_p}{\tau_R}\rangle\langle\frac{\tilde{E}_p}{\tau_R}\tilde{E}_p^2\rangle}\nonumber\\
&-\tilde{p}_{\langle\nu\rangle}\frac{\langle\frac{\tilde{E}_p}{\tau_R}\phi \tilde{p}^{\langle\nu\rangle}\rangle}
{\frac{1}{3}\langle\frac{\tilde{E}_p}{\tau_R}\tilde{p}^{\langle\mu\rangle}\tilde{p}_{\langle\mu\rangle}\rangle}\bigg]~,
\label{MDRTAcoll}
\end{align}
with $\langle\cdots\rangle=\int dF_p(\cdots)$. Eq.\eqref{MDRTAcoll} readily gives ${\cal{L}}_{\text{MDRTA}}[\phi]=0$ if $\phi=a+b(p\cdot u)+c^{\mu}p_{\langle\mu\rangle}$ with $a,b,c^{\mu}$ being arbitrary momentum 
independent coefficients. It satisfies the self adjoint property as well, $\int d\Gamma_p\psi{\cal{L}}_{\text{MDRTA}}[\phi]=\int d\Gamma_p\phi{\cal{L}}_{\text{MDRTA}}[\psi]$. These two combinedly give the summation 
invariant property $\int d\Gamma_p\psi{\cal{L}}_{\text{MDRTA}}[\phi]=0$ for $\psi=a+b(p\cdot u)+c^{\mu}p_{\langle\mu\rangle}$ which immediately results in the conservation laws $\partial_{\mu}N^{\mu}=0$ and 
$\partial_{\mu}T^{\mu\nu}=0$ microscopically. These conservation laws are not needed to be estimated order by order and are treated non-perturbatively. The preservation of particle number and energy-momentum 
conservation in ${\cal{L}}_{\text{MDRTA}}[\phi]$ is irrespective of the frame indices or particular momentum dependence of $\tau_R$. Eq.\eqref{MDRTAcoll} resembles the novel relaxation time collision operator 
introduced in \cite{Rocha:2021zcw} apart from the fact that it uses the polynomial basis given in Eq. \eqref{poly1}-\eqref{poly2}. The advantage of using this basis is that the polynomials associated with the 
homogeneous part of the solution are in form of simple exponents which reduces the computational complexity significantly.

In the current analysis, the momentum dependence of $\tau_R$ is expressed as a power law of $\tilde{E}_{p}$ in the comoving frame, with $\tau_R^0$ as the momentum independent part; the parameter $\Lambda$ 
specify the power of the scaled energy. 

To solve Eq.~\eqref{RTE}, we adopt a perturbative expansion introduced in \cite{Rocha:2022ind}. By decomposing the space-time derivative, the left hand side of Eq.\eqref{RTE} gives rise to a number
of time and space derivatives over the fundamental thermodynamic quantities $T,\mu$ and $u^{\mu}$. In popular perturbation approaches like Chapman-Enskog method, the time derivatives are replaced by the spatial ones
in order to make the left hand side of Eq.\eqref{RTE} orthogonal to zero modes (homogeneous solutions). By the virtue of the collision operator ${\cal{L}}_{\text{MDRTA}}$ given in Eq.\eqref{MDRTAcoll}, the 
right hand side of Eq.~\eqref{RTE} now retains only the interaction part of $\phi$. It singularly excludes the zero modes of the linearized collision operator, i.e, any function proportional to $1$ and $p^{\mu}$ are
not present from the momentum basis of the unknown coefficients in Eq.\eqref{phi1}. Because of the fact, the left hand side of Eq.\eqref{RTE} is not necessarily needed to be orthogonal to zero modes in order extract
the remaining non-zero mode coefficients, which are itself orthogonal to zero modes as well as among themselves. Hence, the covariant time derivatives appearing on the left hand side of Eq.\eqref{RTE} are not 
required to be exchanged by the spatial gradients. Employing that, the inhomogeneous or interaction part of the first-order out-of-equilibrium distribution function turns out to be,
\begin{align}
&\frac{\phi_{\text{int}}^{(1)}}{\tau_R^0}=
-\tilde{E}_{p}^{\Lambda-1}\bigg[\tilde{E}_{p}^2\frac{DT}{T}+\tilde{E}_{p} D\tilde{\mu}+\left(\frac{\tilde{E}_{p}^2}{3}-\frac{z^2}{3}\right)(\partial\cdot u)+\nn
&\tilde{E}_{p}\tilde{p}^{\langle\mu\rangle}\left(\frac{\nabla_{\mu}T}{T}-Du_{\mu}\right)
+\tilde{p}^{\langle\mu\rangle}\nabla_{\mu}\tilde{\mu}-\tilde{p}^{\langle\mu}\tilde{p}^{\nu\rangle}\sigma_{\mu\nu}\bigg],
\label{phiint1}
\end{align}
where $\sigma_{\mu\nu}=\nabla_{\langle{\mu}}u_{\nu\rangle}$, $D=u^{\mu}\partial_{\mu}$, and $\nabla^{\mu}=\Delta^{\mu\nu}\partial_{\nu}$ are symmetric trace-less shear tensor, temporal and spatial counterparts of 
the total space-time derivative respectively. Next, we use Eq.~\eqref{phiint1} in Eq.~\eqref{phi5} to construct $\phi^{(1)}$ in order to calculate the first-order field correction coefficients. 
From Eq.\eqref{rho2}-\eqref{pnflux2}, the first-order thermodynamic field corrections in a general frame and with arbitrary interactions are given by:
\begin{align}
&\delta n^{(1)}, \delta \epsilon^{(1)}, \delta P^{(1)}=\nu_{1},\varepsilon_{1},\pi_{1} \frac{DT}{T} + \nu_{2} ,\varepsilon_{2},\pi_{2}\left(\partial \cdot u\right)\nonumber\\
&~~~~~~~~~~~~~~~~~~~~~~+\nu_{3},\varepsilon_{3},\pi_{3} D\tilde{\mu}~,
\label{sc}\\
&W^{(1)\mu},V^{(1)\mu} = \theta_{1},\gamma_{1}\left[\frac{\nabla^{\mu}T}{T} - Du^{\mu}\right] +\theta_{3},\gamma_{3} \nabla^{\mu}\tilde{\mu}.
\label{vec}
\end{align}
The explicit expressions of the field correction coefficients turn out to be elaborate and complicated functions of the frame indices $i,j,k$ and the parameter $\Lambda$ of MDRTA. These field corrections along with 
$\pi^{(1)\mu\nu}=2\eta\sigma^{\mu\nu}$ ($\eta$ is shear viscosity), constitute the first order out of equilibrium $N^{\mu}$ and $T^{\mu\nu}$ from Eq.~\eqref{numberflow} and Eq.~\eqref{enmomflow}, respectively. 

So, here we end up with 14 field correction coefficients ($\nu_{1,2,3},\epsilon_{1,2,3},\pi_{1,2,3},\theta_{1,3},\gamma_{1,3}$ and $\eta$). It was shown in~\cite{Kovtun:2019hdm,Hoult:2021gnb} 
that not all coefficients are invariant under the first-order field redefinition 
(due to the arbitrariness in the definition of temperature, fluid four-velocity and chemical potential for out-of equilibrium case). We checked that our coefficients satisfy the combinations
$f_i=\pi_i-\varepsilon_i\left(\frac{\partial P_0}{\partial \epsilon_0}\right)_{n_0}-\nu_i \left(\frac{\partial P_0}{\partial n_0}\right)_{\epsilon_0}$ and 
$l_i=\gamma_i-\frac{n_0}{\epsilon_0+P_0}\theta_i$ to be frame invariant (i.e., independent of the indices $i,j,k$), which further reduce to the physical transport coefficients; bulk viscosity 
$\zeta=-f_2+\left(\frac{\partial P_0}{\partial \epsilon_0}\right)_{n_0}f_1+\frac{1}{T}\left(\frac{\partial P_0}{\partial n_0}\right)_{\epsilon_0}f_3$, and charge conductivity 
$k_n=l_3-\frac{n_0T}{(\epsilon_0+P_0)}l_1$. The detailed expressions of $\zeta$ and $k_n$ with MDRTA are given in~\cite{Mitra:2021owk}. The corrections further reveal that, the LL and Eckart limit of the scalar 
indices ($i=1,j=2$ or vice versa) give $\delta n^{(1)}=0,\delta \epsilon^{(1)}=0$ (such that $\zeta$ is entirety taken up by the pressure correction), where for the vector index, LL limit ($k=1$) gives $W^{(1)\mu}=0$ 
and Eckart limit ($k=0$) gives $V^{(1)\mu}=0$. Most significantly, we found that for the momentum independent relaxation time (i.e., for $\Lambda$=0), all the correction coefficients associated with the first-order 
time derivatives $(\nu_1,\nu_3,\varepsilon_1,\varepsilon_3,\pi_1,\pi_3,\theta_1,\gamma_1)$ in Eqs.~\eqref{sc}-\eqref{vec}  identically vanish for all hydrodynamic frame conditions (irrespective of $i,j,k$ values) 
which will be shown later to have crucial implications on the causality and stability of the theory.

\textit{Stability and causality analysis.}$\textendash$
Here we investigate the causality and stability of the theory by linearizing the conservation equations for small perturbations of fluid variables around the hydrostatic equilibrium in the local rest frame, $\epsilon(t,x)=\epsilon_0+\delta\epsilon(t,x),~n=n_0+\delta n(t,x),,P(t,x)=P_0+\delta P(t,x),~u^{\mu}(t,x)=(1,\vec{0})+\delta u^{\mu}(t,x)$. In linear approximation, $\delta u^{\mu}$ has only spatial components to retain the normalization condition. For convenience, these fluctuations are further expressed in their plane wave solutions via a Fourier 
transformation $\delta\psi(t,x)\rightarrow e^{i(\omega t-kx)} \delta\psi(\omega,k)$, with wave 4-vector $k^{\mu}=(\omega,k,0,0)$. The resulting dispersion relation for transverse or shear channel is,
\be
(i\omega)^2+i\omega\frac{(\epsilon_0+P_0)}{\theta}+\frac{\eta}{\theta}k^2=0~,
\label{dispshear}
\ee
where we define $\theta=-\theta_1$. At small $k$ limit, the obtained modes are, $\omega^{T}_1=i\frac{\eta}{(\epsilon_0+P_0)}k^2+{\cal{O}}(k^4)$ and $\omega^T_2=i\frac{(\epsilon_0+P_0)}{\theta}+{\cal{O}}(k^2)$.
Both the modes are non-propagating, where $\omega^{T}_1$ is a hydrodynamic mode (vanishes at $k=0$) and $\omega^{T}_2$ is a non-hydro mode. $\omega^{T}_1$ is the conventional shear mode of NS theory. 
At small $k$ the stability is guaranteed if $\theta>0$, because in that case the imaginary part of $\omega_2^{T}$ is positive definite and gives rise to exponentially decaying perturbations.
At large $k$, the modes come out to be $\omega^{T}_{1,2}=\pm \sqrt{\eta/\theta}~k+i\frac{(\epsilon_0+P_0)}{2\theta}+{\cal{O}}(\frac{1}{k})$.
These are propagating modes where causality holds for $\theta>\eta$, which also guarantees the stability condition. $\theta_1$ plays a crucial role in stability and causality of the shear channel. 
From Eq.~\eqref{vec} the explicit expression of $\theta_1$ turns out to be,
\be
\theta_1=-\tau_R^0 T^2 \left( J_{\Lambda+1}+\frac{\epsilon_0+P_0}{T^2}\frac{J_{k+\Lambda}}{J_k}\right)~.
\ee
\begin{figure}[h]
    \centering
    \includegraphics[width=0.48\textwidth]{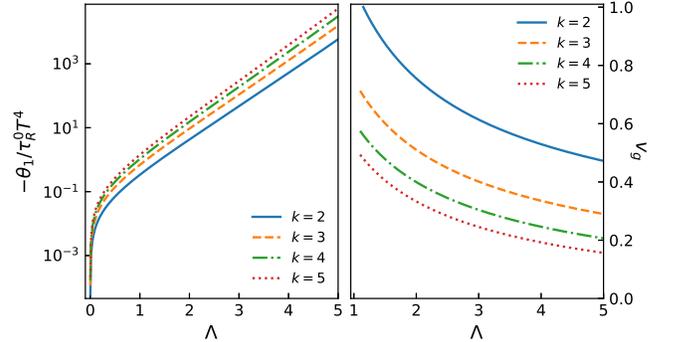}
    \caption{$\theta_1$ and $v_g$ as a function of $\Lambda$ in general frames}
    \label{theta1}
\end{figure}
We can see, $\theta_1=0$ for both $k=1$ (LL frame) with any interaction or $\Lambda=0$ (momentum independent RTA) for any general frame. This will give rise to superluminal velocities in the shear 
channel. In Fig.~\eqref{theta1}, left panel shows $\theta(=-\theta_1)$ scaled by $\tau_R^0$ as a function of $\Lambda$ for different vector matching indices $k$. $\theta$ is always positive for 
$\Lambda>0$. The right panel shows the group velocity $v_g=\sqrt{\eta/\theta}$ which obeys causality for $\Lambda>1$, where $\eta=\tau_R^0T^2K_{\Lambda-1}/2$ with 
$\Delta^{\alpha\beta\mu\nu}K_n=\int dF_p\tilde{p}^{\langle\mu}\tilde{p}^{\nu\rangle}\tilde{p}^{\langle\alpha}\tilde{p}^{\beta\rangle}\tilde{E}_{p}^n~$. We also see that, larger values of $k$ and $\Lambda$
reduce group velocity. So,  even in a general frame, the choice of $\Lambda$ crucially decides the stability and causality of the shear channel. Throughout the numerical analysis, the parameters have been set to, 
$T=300$ MeV, $m=300$ MeV.

For longitudinal or sound mode, the dispersion relation turns out to be a sixth order polynomial,
\begin{align}
&(i\omega)^6A_6+(i\omega)^5A_5+(i\omega)^4A_4+(i\omega)^3A_3\nn
&+(i\omega)^2A_2+(i\omega)A_1+A_0=0~,
\label{dispsound}
\end{align}
with $A_4=A_4^0+A_4^2k^2, A_3=A_3^0+A_3^2k^2, A_2=A_2^2k^2+A_2^4k^4, A_1=A_1^2k^2+A_1^4k^4, A_0=A_0^4k^4+A_0^6k^6$. 
Eq.~\eqref{dispsound} agrees with result obtained in~\cite{Taghinavaz:2020axp}, where the coefficient $A$'s are functions of $\nu_{1,2,3},\varepsilon_{1,2,3},\pi_{1,2,3},\theta_{1,3},\gamma_{1,3}$ defined earlier 
(the detailed analysis will be reported elsewhere). Eq.~\eqref{dispsound} cannot be solved analytically and hence we present results for $k\rightarrow 0$ limit. At this limit,
Eq.~\eqref{dispsound} gives three hydrodynamic modes as, $\omega_6^L=i\hat{h}^2\frac{k_nT}{(\epsilon_0+P_0)}k^2$ and $\omega_{4,5}^L= \pm c_s k+i\frac{\Gamma_s}{2} k^2+{\cal{O}}(k^3)$, 
with scaled enthalpy per particle $\hat{h}=(\epsilon_0+P_0)/n_0T$, 
velocity of sound squared $c_s^2=\big(\frac{\partial P_0}{\partial\epsilon_0}\big)_{n_0}+\frac{1}{\hat{h}}\frac{1}{T}\big(\frac{\partial P_0}{\partial n_0}\big)_{\epsilon_0}$
and sound attenuation coefficients $\Gamma_s=\left[\frac{4}{3}\eta+\zeta+\frac{k_nT}{c_s^2}\big(\frac{1}{T}\frac{\partial P_0}{\partial n_0}\big)^2_{\epsilon_0}\right]/(\epsilon_0+P_0)$.
$\omega_6^L$ and $\omega_{4,5}^L$ are the conventional heat-diffusion and sound modes of the NS theory respectively.

The remaining non-hydro modes are given by, 
\be
(i\omega^L)^3A_6+(i\omega^L)^2A_5+(i\omega^L)A^0_4+A^0_3=0~.
\label{stab1}
\ee
Using Routh-Hurwitz criteria, we find the following conditions for stability of the non-hydro modes,
\begin{align}
&A_6>0~,~A_5>0~,~A_3^0>0~,
\label{stab2}\\
&B_2=(A_4^0A_5-A_3^0A_6)/A_5>0~.
\label{stab3}
\end{align}
Among these coefficients, $A_3^0=n_0(\epsilon_0+P_0)$ is always positive. The remaining coefficients are given by,
\begin{align}
 A_6&=\frac{\theta_1}{(\epsilon_0+P_0)}\hat{h}c_s^2(\nu_1\epsilon_3-\nu_3\epsilon_1)~,\\
 A_5&=\hat{h}c_s^2(\nu_3\epsilon_1-\nu_1\epsilon_3)\nn
 &-\theta_1\left[(\nu_1 f+\nu_3 c)+\frac{1}{\hat{h}T}(\epsilon_1 g+\epsilon_3 d)\right]~,\\
 A_4^0&=(\epsilon_0+P_0)(\nu_1 f+\nu_3 c)+n_0(\epsilon_1 c+\epsilon_3 d-\theta_1)~.
\end{align}
with, 
$c=J_0 I_3/(I_2^2 - I_1 I_3)~,~d=-J_1 I_2/(I_2^2 - I_1 I_3)~,~f=-J_0 I_2/(I_2^2 - I_1 I_3)~,~g=J_1 I_1/(I_2^2 - I_1 I_3)$. 
\begin{figure}
\includegraphics[width=0.35\textwidth]{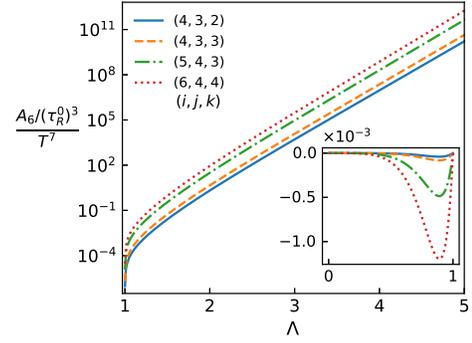}
\caption{$A_6$ as a function of $\Lambda$ in general frames}
\label{A6}
\end{figure}
The concerned field correction coefficients are given by,
\begin{align}
&\varepsilon_1=\tau_R^0\left[\frac{\partial\epsilon_0}{\partial\tilde{\mu}}\frac{{\cal{D}}_{i,j}^{\Lambda+1,1}}{{\cal{D}}_{i,j}^{0,1}}+
T\frac{\partial\epsilon_0}{\partial T}\frac{{\cal{D}}_{i,j}^{\Lambda+1,0}}{{\cal{D}}_{i,j}^{1,0}}-T^2I_{\Lambda+3}\right],\\
&\varepsilon_3=\tau_R^0\left[\frac{\partial\epsilon_0}{\partial\tilde{\mu}}\frac{{\cal{D}}_{i,j}^{\Lambda,1}}{{\cal{D}}_{i,j}^{0,1}}+
T\frac{\partial\epsilon_0}{\partial T}\frac{{\cal{D}}_{i,j}^{\Lambda,0}}{{\cal{D}}_{i,j}^{1,0}}-T^2I_{\Lambda+2}\right]~,\\
&\nu_1=\tau_R^0\left[\frac{\partial n_0}{\partial\tilde{\mu}}\frac{{\cal{D}}_{i,j}^{\Lambda+1,1}}{{\cal{D}}_{i,j}^{0,1}}+
T\frac{\partial n_0}{\partial T}\frac{{\cal{D}}_{i,j}^{\Lambda+1,0}}{{\cal{D}}_{i,j}^{1,0}}-TI_{\Lambda+2}\right]~,\\
&\nu_3=\tau_R^0\left[\frac{\partial n_0}{\partial\tilde{\mu}}\frac{{\cal{D}}_{i,j}^{\Lambda,1}}{{\cal{D}}_{i,j}^{0,1}}+
T\frac{\partial n_0}{\partial T}\frac{{\cal{D}}_{i,j}^{\Lambda,0}}{{\cal{D}}_{i,j}^{1,0}}-TI_{\Lambda+1}\right]~.
\end{align}
The coefficients $\nu_1,\varepsilon_1$ vanish both for 
$i=1, j=2$  (LL+Eckart) $\forall \Lambda$ and also at $\Lambda=0$ for all frame choices. $\nu_3,\varepsilon_3$ obey the same but also vanish for $\Lambda=1$ at all frames. The coefficients make $A_5$ and $A_4^0$ 
vanish for $\Lambda=0$ and $A_6$ vanish for both $\Lambda=0$ and $1$ at any frame.
\begin{figure}
\includegraphics[width=0.5\textwidth]{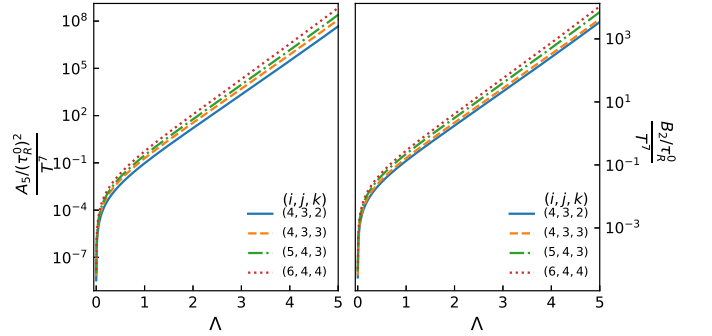}
\caption{$A_5$ and $B_2$ as a function of $\Lambda$ in general frames}
\label{A5B2}
\end{figure}
From Fig.~\eqref{A6} we can see that $A_6$ becomes positive for $\Lambda>1$, but it becomes negative for the region $\Lambda$= 0 to 1 excluding the end points.
This is shown in the inset of Fig.~\eqref{A6}, where we can see that in this region higher values of the frame indices make the situation worse with larger negative values of $A_6$, 
resulting in more increased instability. This is the essence of the current work. In short, we conclude that a general frame and 
the nature of underlying interactions are both crucial for the stability and causality of a first-order theory.
Fig.~\eqref{A5B2} shows the dependency of $A_5$ and $B_2$ on $\Lambda$ for different frames, which turns out to be positive for general frames and $\Lambda>0$. 

\textit{Conclusion.}
$\textendash$ In this work, a first-order, relativistic stable and causal hydrodynamic theory has been derived in a general frame from the Boltzmann transport equation, where the system interactions are introduced 
via the microscopic particle momenta captured through $\tau_R$ and an appropriate collision operator ${\cal{L}}_{\text{MDRTA}}$. We have shown that in order to hold stability and causality at 
first-order theories, besides a general frame, the system interactions need to be carefully taken into account. The conventional momentum independent RTA leads to acausality by diverging the shear modes even in a 
general frame. The momentum dependence employed through MDRTA is shown to subtly control the stability and causality of the theory in a general frame.

We believe that this correlation between system dynamics (microscopic interactions) and relativistic hydrodynamics (macroscopic frame variables), along with the precise estimation of causality and stability 
conditions, makes the current work an acceptable first-order hydrodynamic theory, ready for practical applications. 

\textit{Acknowledgements.}$\textendash$ 
R.B. and V.R. acknowledge financial support from the DST Inspire faculty research grant (IFA-16-PH-167), India. S.M. acknowledges funding support from DNAP, TIFR, India.

\end{document}